\documentstyle{ichep}
\newcommand{\be}{\begin{equation}}
\newcommand{\ee}{\end{equation}}
\newcommand{\bea}{\begin{eqnarray}}
\newcommand{\eea}{\end{eqnarray}}
\newcommand{\bel}{\begin{leqnarray}}
\newcommand{\eel}{\end{leqnarray}}
\newcommand{\bc}{\begin{center}}
\newcommand{\ec}{\end{center}}
\newcommand{\nn}{\nonumber}

\newcommand{\ms}{m_s(\mu)}

\newcommand{\ep}{\epsilon}

\newcommand{\epse}{\epsilon^\prime/\epsilon}

\newcommand{\DBtwo}{\Delta B\!=\!2}
\newcommand{\DSone}{\Delta S\!=\!1}

\newcommand{\Gmulup}{\gamma^\mu_L}

\begin{document}

\title{Next-to-leading prediction of $\epse$: an upgraded analysis}

\author{Marco Ciuchini$^{\dag}$}

\affil{INFN, Sezione Sanit\`a, \\
V.le Regina Elena 299, 00161 Roma, Italy}

\abstract{We present an updated theoretical prediction of $\epse$, using the
next-to-leading $\DSone$ effective hamiltonian and lattice QCD matrix elements.
The CP violating phase is costrained by using both the experimental values of
$\ep$
and $x_d$, assuming the theoretical determination of $f_B$.
Predictions of $\cos\delta$ and $\sin 2\beta$ are also obtained in this way.
For $\epse$, our estimate is $\epse=(2.8 \pm 2.4)\times 10^{-4}$.}

\twocolumn[\maketitle]

\fnm{1}{E-mail: ciuchini@vaxsan.iss.infn.it.}

We repeat the combined analysis of the CP violation parameter $\ep$ and the
$B$-mixing parameter $x_d$ in order to estimate $\epse$, along the lines
followed
in refs. \cite{reina,ciuc1}. The main steps of this analysis are the following:
\begin{enumerate}
\item The CP violating phase $\delta$ of the CKM matrix is constrained by
comparing the theoretical prediction for $\ep$ with its experimental value.
To this purpose, the relevant formula is
\bea
\label{epsilon}
|\epsilon|_{\xi=0}&=&C_\epsilon B_KA^2\lambda^6\sigma\sin\delta
\left\{F(x_c,x_t)+\right.\\
& & \left.F(x_t)[A^2\lambda^4(1-\sigma\cos\delta)]-F(x_c)\right\},\nn
\eea
where $x_q={m_q^2}/{M_W^2}$ and
the functions $F(x_i)$ and $F(x_i,x_j)$ are the so-called
{\it Inami-Lim} functions \cite{inami}, obtained from the calculation of the
basic box-diagram and including QCD corrections. $F(x_t)$ is known at the
next-to-leading order, which has been included in our calculation \cite{bur0}.
In eq. \eref{epsilon},
\be
C_\epsilon=\frac{G_F^2f_K^2 m_ KM_W^2}{6\sqrt 2\pi^2\Delta M_K},
\ee
where $\Delta M_K$ is the mass difference between the two neutral kaon mass
eigenstates.
Moreover, $\rho=\sigma\cos\delta$ and $\eta=\sigma \sin\delta$, where
$\lambda$, $A$,
$\rho$ and $\eta$ are the parameters of the CKM matrix in the Wolfenstein
parametrization \cite{wolf}. Finally, $B_K$ is the renormalization group
invariant
$B$-factor \cite{slom}.

\item The theoretical estimate of the $B$-meson coupling constant
$f_B$ is used to further constrain $\delta$, by comparing the theoretical
prediction
of the $B$ mixing parameter $x_d$ with its experimental value. From the
$\DBtwo$ effective hamiltonian, one can derive
\bea
x_d &=&\frac{\Delta M}{\Gamma}=C_B \frac{\tau_B f_B^2}{M_B} B_B
A^2 \lambda^6 \Bigl( 1 +\nn\\
& & \qquad\qquad \sigma^2 - 2 \sigma cos\delta \Bigr) F(x_t),\\
C_B &=& \frac{G_F^2 M_W^2 M_B^2}{6 \pi^2},\nn
\eea
where $B_B$ is the $B$-parameter relevant for $B-\bar B$ matrix element
\be
\langle\bar B_d|(\bar d\Gmulup b)^2|B_d\rangle= 8/3 f_B^2 M_B^2 B_B.
\ee
Notice that $f_B B_B^{1/2}$ must be known, for the experimental value
of $x_d$ to give a constraint on $\delta$.

\item From the $\DSone$ effective hamiltonian, one can calculate the expression
of $\epsilon^\prime$ in terms of CKM matrix elements, Wilson coefficients and
local
operator matrix elements. One has
\be
\epsilon^{\prime}=\frac{e^{ i\pi/4}}{\sqrt{2}}\frac{\omega}
{\mbox{Re}A_{ 0}}\left[\omega^{ -1}
(\mbox{Im}A_{ 2})^{\prime}-(1-\Omega_{ IB})\,\mbox{Im}A_{ 0}\right],
\ee
where
$(\mbox{Im}A_2)^\prime$ and $\mbox{Im}A_0$ are given by
\bea
\label{ima0}
\mbox{Im}A_0 &=&-G_F Im\Bigl({ V}_{ts}^{*}V_{td}\Bigr)
\left\{-\left(C_6 B_6+\right.\right.\nn\\
& & \left.\frac{1}{3}C_5 B_5\right)Z+
\left(C_4 B_4+\frac{1}{3}C_3 B_3\right)X+\nn\\
& &C_7B_7^{1/2}\left(\frac{2Y}{3}+\frac{Z}{6}+\frac{X}{2}\right)+\\
& & C_8 B_8^{1/2}\left(2Y+\frac{Z}{2}+\frac{X}{6}\right)-\nn\\
& &\left.C_9B_9^{1/2}\frac{X}{3}+\left(\frac{C_1
B_1^c}{3}+C_2 B_2^c\right)X\right\},\nn
\eea
and
\bea
\label{ima2}
(\mbox{Im}A_2)^{\prime}&=&-G_FIm\Bigl(V_{ts}^{*}V_{td}\Bigr)
\left\{C_7 B_7^{3/2}\left(\frac{Y}{3}-\right.\right.\nn\\
& & \left.\frac{X}{2}\right)+
C_8 B_8^{3/2}\left(Y-\frac{X}{6}\right)+\\
& & \left.C_9 B_9^{3/2}\frac{2X}{3}\right\}.\nn
\eea
Here $\omega=\mbox{Re}A_2/\mbox{Re}A_0$ and
we have introduced $(\mbox{Im}A_2)^\prime$ defined as
\be
\mbox{Im}A_2=(\mbox{Im}A_2)^\prime+\Omega_{IB}(\omega\mbox{Im}A_0).
\ee
$\Omega_{IB}$ accounts for the isospin breaking contribution, see for example
ref.
\cite{bur5}.
The Wilson coefficients $C_i$ have been
evaluated at the next-to-leading order for $\mu=2$ GeV, using the anomalous
dimension matrices given in refs. \cite{bur3,ciuc2} and the initial conditions
computed in refs. \cite{flynn1,bur1} (given for HV in ref. \cite{bur2}).
Concerning the local operator matrix elements, their values are given by a set
$\{B_{ i}\}$ of $B$-parameters multiplied by the vacuum insertion
approximation results. In turn, these can be
written in terms of the three quantities (see eq. \eref{ima0} and
eq. \eref{ima2})
\bea
X &=& f_{\pi}\left(M_{ K}^{ 2}-M_{\pi}^{ 2}\right), \nn\\
Y &=& f_{\pi}\left(\frac{M_{ K}^{ 2}}{m_s(\mu)+m_d(\mu)}\right)^2\nn\\
&\sim& 12\,X\left(\frac{0.15 \, \mbox{GeV}}{m_s(\mu)}\right)^2, \\
Z &=& 4\left(\frac{f_{ K}}{f_{\pi}}-1\right)Y.\nn
\eea
The numerical values of the $B$-parameters have been taken from lattice
calculations \cite{lattice}
For those $B$-factors which have not yet been computed on the lattice,
we have used educated guesses, see ref. \cite{ciuc3}.
\end{enumerate}

More details on this analysis can be found in ref.
\cite{ciuc3,vietnam}.
Compared to our previous work \cite{ciuc1}, the main improvements are
the following:
\begin{enumerate}
\item The constraint on $\delta$ coming from $x_d$ is used in
the analysis, taking $f_B$ from the theory.
Since there is increasing theoretical evidence that the value
of $f_B$ is large ($\sim 200$ MeV) and that the relevant $B$-parameter $B_B$ is
close to one, this constraint is quite effective.
\item Updated  values of the experimental parameters entering in
the phenomenological analysis, such as the B meson lifetime $\tau_B$,
the $B^0_d$--$\bar B^0_d$ mixing parameter $x_d$,
the CKM matrix elements, ($\vert V_{cb} \vert$,
$\vert V_{ub}\vert/\vert V_{cb}\vert $),  etc., have been used.
\item The value of the strange quark mass $m_s$ has been taken from
lattice calculations \cite{ms}, thus making a more consistent use of lattice
results for
the B-parameters of the relevant penguin operators.
\item All the results are presented with an estimate of the corresponding
errors. These errors come from the limited precision of measured quantities,
e.g. $\tau_B$, and from theoretical uncertainties, e.g. the
values of hadronic matrix elements.
\end{enumerate}

\begin{table}
\Table{|c|c|}{
\hline
\multicolumn{2}{|c|}{Parameters}\\ \hline
$(f_B B_B^{1/2})_{th}=(200\pm 40)$ MeV &
$V_{cb}=A\lambda^2=0.040\pm 0.006$ \\
$m_s(2 \mbox{GeV})=(128\pm 18)$ MeV & $x_d=0.685\pm 0.076$ \\
$\Lambda_{QCD}^{n_f=4}=(330\pm 100)$ MeV & $\Omega_{IB}=0.25\pm 0.10$ \\
$\tau_B =(1.49\pm 0.12)\times 10^{-12}$ s & $m_t=(174 \pm 17)$ GeV \\
$\vert V_{ub}/V_{cb}\vert=\lambda\sigma=0.080\pm 0.015$ &  \\
\hline
}
\caption{Values of the fluctuating parameters used in the numerical analysis.}
\label{tab:par}
\end{table}

The results of our analysis have been obtained by varying the experimental
quantities, e.g. the value of the top mass $m_t$, $\tau_B$,  etc. and
the theoretical parameters, e.g. the  B-parameters,
the strange quark mass $\ms$, etc., according to their errors.
Values and errors of the input quantities used in the following are reported
in tables \ref{tab:par}--\ref{tab:bpar}.
We assume a gaussian distribution
for the experimental quantities and a flat distribution
(with a width of 2$\sigma$) for the theoretical ones.
The only exception is $\ms$, taken from quenched lattice $QCD$ calculations,
for which we have assumed a gaussian distribution, according to the results
of ref. \cite{ms}.
\par
The theoretical predictions ($\cos \delta$, $\sin 2\beta$, $\epse$, etc.)
depend on several fluctuating parameters. We have obtained numerically their
distributions, from which we have calculated the central values and the errors
reported below.
\begin{table}
\Table{|c|c|}{
\hline
\multicolumn{2}{|c|}{Constants}\\ \hline
$G_F=1.16634\times 10^{-5}\mbox{GeV}^{-2}$ & $f_{\pi}=132$ MeV \\
$m_c=1.5$ GeV & $f_K=160$ MeV \\
$m_b=4.5$ GeV & $\lambda=\sin\theta_c=0.221$ \\
$M_W=80.6$ GeV & $\ep_{exp}=2.268\times 10^{-3}$ \\
$M_{\pi}=140$ MeV & $\mbox{Re}A_0=2.7\times 10^{-7}$ GeV \\
$M_K=490$ MeV & $\omega=0.045$ \\
$M_B=5.278$ GeV & $\mu=2$ GeV\\
$\Delta M_K=3.521\times 10^{-12}$ MeV & \\
\hline
}
\caption{Values of the constants used in the numerical analysis.}
\label{tab:cons}
\end{table}

\widefigure{23pc}{Distributions of values for $\cos \delta$,
$\sin 2\beta$ and $\epse$, using the values of the parameters given in the
tables. The solid istograms are obtained without using
the $x_d$ constraint. The dashed ones use
this constraint, assuming that $f_B B_B^{1/2}=200\pm 40$ MeV.
The contour-plot of the event distribution in the $\rho-\eta$ plane is also
shown.\label{fig:cd}}

Using the values given in the tables and the formulae given previously, we have
obtained the following results:
\begin{enumerate}
\item
The distribution for $\cos \delta$, obtained by comparing
the experimental value of $\ep$ to its theoretical
prediction, is given in \fref{fig:cd}. As already noticed in
refs. \cite{reina,ciuc1} and \cite{schubert,alig},
large values of $f_B$ and $m_t$ favour $\cos \delta > 0$,
given the current measurement of $x_d$. When the condition $160$ MeV
$\le f_B B_B^{1/2}\le 240$ MeV is imposed ($f_B$-cut),
most of the negative solutions disappear,
giving the dashed istogram of \fref{fig:cd}, from which we estimate
\be
\cos \delta= 0.47 \pm 0.32\,\, .
\ee
\item The value of $\sin 2 \beta$ depends on
$\cos \delta$.  The distribution of $\sin 2 \beta$ is shown in
\fref{fig:cd}, without (solid)  and with (dashed) the $f_B$-cut.
When the $f_B$-cut is imposed,
one gets larger values of $\sin 2 \beta$ \cite{reina}.
{}From the dashed distribution, we obtain
\be \sin 2 \beta=0.65 \pm 0.12\, \, .
\ee
\Fref{fig:cd} also contains the contour-plot of the event distribution in the
$\rho-\eta$ plane, showing the effect of the $\ep$ and $x_d$ constraints,
when the $f_B$-cut is imposed.

\item In \fref{fig:tuttoepe}, several informations
on $\epse$ are provided. Contour-plots of the distribution of the generated
events in the $\epse$--$\cos\delta$ plane are shown, without
and with the $f_B$-cut. One notices a very mild dependence
of $\epse$ on $\cos \delta$. As a consequence one obtains approximatively
the same prediction in the two cases (see also \fref{fig:cd})
\be
\epse = (2.3 \pm 2.1 )\times  10^{-4} \,\,\, {\rm no-cut},
\ee
and
\be
\epse = (2.8 \pm 2.4 )\times  10^{-4} \,\,\, f_B-{\rm cut}.
\ee
In \fref{fig:tuttoepe}, we also give $\epse$ as a function of $m_t$. The
band corresponds to the $2\sigma$ prediction.
\end{enumerate}

\begin{table}
\Table{|c|c|}{
\hline
\multicolumn{2}{|c|}{$B$-parameters}\\ \hline
$B_K=0.75\pm 0.15$ & $B_9^{(3/2)}=0.62\pm 0.10$ \\
$B_{1-2}^c=0-0.15^{(*)}$ & $B_{3,4}=1-6^{(*)}$ \\
$B_{5,6}=B_{7-8}^{(3/2)}=1.0\pm 0.2$ & $B_{7-8-9}^{(1/2)}=1^{(*)}$\\
\hline
}
\caption{Values of the $B$-parameters, for operators renormalized at the
scale $\mu=2$ GeV. The only exception is $B_K$, which is the RG
invariant $B$-parameter. $B_9^{3/2}$ has been taken equal to $B_K$, at any
scale. The value reported
in the table is $B_9^{3/2}(\mu=2 {\rm GeV})$.
Entries with a $^{(*)}$ are educated guesses, the others are taken from lattice
QCD calculations.}
\label{tab:bpar}
\end{table}

In spite of several differences,
the bulk of our results overlap with those of ref. \cite{burasepe}.
It is reassuring that theoretical predictions,
obtained by using different approaches to evaluate
the operator matrix elements, are in good agreement.

\widefigure{24pc}{Above, contour-plots of the event distributions in the
plane $\epse$--$\cos \delta$ without and with the $f_B$-cut. Below,
$\epse$ as a function of $m_t$.\label{fig:tuttoepe}}
On the basis of the latest analyses, it seems very difficult
for $\epse$ to be larger than $10 \times 10^{-4}$.
This may happen by taking the matrix elements of the
dominant operators, $Q_6$ and $Q_8$, much different than usually assumed.
One possibility, discussed in ref. \cite{burasepe}, is to take
$B_6 \sim 2$ and $B_8 \sim 1$, instead of the usual values $B_6 \sim B_8 \sim
1$.
To our knowledge, no coherent theoretical approach can accomodate
so large value of $B_6$.
\section*{Acknowledgments}
The precious collaboration of E. Franco, G. Martinelli and L. Reina is
acknowledged.

\end{document}